\documentclass[reprint,titlepage,amsmath,amssymb,aps,physrev,floatfix,showkeys,nofootinbib]{revtex4-2}

\usepackage{graphicx}
\usepackage[utf8]{inputenc}
\usepackage[T1]{fontenc}
\usepackage{newtxtext}
\usepackage{newtxmath}
\usepackage{microtype} 
\usepackage{physics} 
\usepackage{siunitx} 
    \DeclareSIUnit{\solarmass}{\ensuremath{M_{\odot}}}
\usepackage[svgnames]{xcolor} 
    \definecolor{accent}{HTML}{df2d16}
\usepackage{tikz}
\usepackage{pgfplots} 
    \pgfplotsset{compat=1.18}
    \usepgfplotslibrary{fillbetween} 
\usepackage[allcolors=accent,colorlinks,pdfusetitle,pdfauthor={Caio César Rodrigues Evangelista and Níckolas de Aguiar Alves}]{hyperref} 
\usepackage{orcidlink} 

\begin{document}
\title{Using thermodynamics to learn gravitational wave physics}

\author{Caio César Rodrigues Evangelista\,\orcidlink{0009-0000-9275-0324}}
\email[Corresponding author: ]{caio@fisica.ufc.br}
\affiliation{Physics Department, Block 922, Campus do Pici, \href{https://ror.org/03srtnf24}{Federal University of Ceará}, Fortaleza, Ceará 60455-760, Brazil}

\author{\firstname{Níckolas} de \surname{Aguiar Alves}\,\orcidlink{0000-0002-0309-735X}}
\email[Contact author: ]{alves.nickolas@ufabc.edu.br}
\affiliation{Center for Natural and Human Sciences, \href{https://ror.org/028kg9j04}{Federal University of ABC}, Avenida dos Estados 5001, Bangú, Santo André, São Paulo 09280-560, Brazil}

\date{09 March 2026}

\begin{abstract}
Black holes are some of the most interesting objects in the universe. While they first arise in the complicated behavior of general relativity, the physical laws ruling their behavior are surprisingly simple. For example, one of the core facts about black holes is that their area never decreases, much like the entropy in thermodynamics. In this note directed at introductory physics students and their instructors, we use this similarity to understand properties of black hole physics using standard techniques from an undergraduate course in thermal physics. We explore the never-decreasing nature of black hole area to obtain bounds on the energy emitted in a black hole merger (a calculation originally done by Hawking). We show how this allows us to think of black holes in manners very similar to heat engines, and how these ideas have been used in modern gravitational wave observatories to test general relativity. This allows a research-level topic to be discussed in introductory physics lectures.
\end{abstract}

\maketitle

\section{Introduction}
    General relativity is one of the most triumphant theories in physics. All gravitational phenomena we ever encountered seem be well-described by its framework, and it still achieves new experimental successes over a century after its theoretical proposal. With each new test, our confidence in the theory is strengthened, and our curiosity about whether it will ever crumble grows. 

    For most of its history, the experimental tests of general relativity were focused on small deviations from Newtonian calculations. For example, the ``classical tests'' consisted in measuring deviations in the perihelion precession of Mercury, the deflection of light by massive bodies, and the gravitational redshift of light \cite{carroll2019SpacetimeGeometryIntroduction}. Nowadays, many new tests of gravity are available in a variety of scenarios. In particular, gravitational wave observatories such as LIGO, Virgo, and KAGRA have allowed us to probe the behavior of very strong gravitational fields. This opens the exciting possibility of finding new gravitational physics beyond general relativity.

    One of the most interesting predictions that has been recently tested through gravitational wave observations is the ``area theorem'' \cite{isi2021TestingBlackHoleArea,tang2026VerificationBlackHole,abac2025GW250114TestingHawkings}. This result by \textcite{hawking1971GravitationalRadiationColliding,hawking1972BlackHolesGeneral} states that, under mild hypotheses, the total area of black holes in the universe never decreases. Hence, the black hole area behaves very similarly to the entropy of usual thermal systems, and we can learn aspects of black hole physics through a straightforward analogy with textbook thermodynamics. In fact, this analogy has now been recognized as a profound statement about the behavior of gravity and developed into the field of black hole thermodynamics. Studying it has taught us much about the inner workings of gravity, thermodynamics, and quantum theory. Our goal in this note is to explain how these ideas can emerge in black hole mergers, and how they have been used in modern gravitational wave observatories to test general relativity in the presence of strong gravitational fields. 

    In the following, we summarize and adapt calculations originally done by \textcite{hawking1971GravitationalRadiationColliding,hawking1972BlackHolesGeneral}. In doing so, we also emphasize the similarities with thermodynamics---which would be uncovered a few years later. The discussion is thus suitable for introductory physics lectures, for instance as a special topic in a first course in thermodynamics.

\section{What Is a Black Hole?}
    Some of the most exciting predictions of general relativity concern the fate of stars. Typical stars are kept in equilibrium through the nuclear fusion reactions in their cores, which lead to huge pressures in their interior. The pressure sustains the star against its own gravitational field, allowing equilibrium. Nevertheless, fuel is finite, and when it inevitably runs out the star will start to collapse under its own weight.

    The fate of a star post-collapse depends hugely on its mass. If the star is heavy enough, the gravitational pull will cause complete collapse. The stellar core will be compacted to the point that no matter or energy is able to escape from it. In particular, not even light is able to leave this region of spacetime. For this reason, we call it a black hole. Its surface is known as its event horizon.

    While the initial star can be extremely complicated, a surprising outcome of general relativity is that black holes are very simple. A result known as the no-hair theorem ensures that, within certain conditions, a stationary black hole is determined by only a handful of parameters---namely, its mass \(M\), its electric charge \(Q\), and its angular momentum (or ``spin'') \(J\). By ``stationary'' we mean that the black hole has settled down to an equilibrium state and its structure no longer changes with time. While the black hole spin is a vector, when dealing with one black hole at a time we can always pick coordinates such that this vector lies along the \(z\)-axis. Hence, the number \(J \geq 0\) given above denotes only the magnitude of the spin.

    Notice there is a stark contrast between the physics of the initial star and of the resulting black hole. Stars can have multiple defining parameters in addition to \(M\), \(Q\), and \(J\). For instance, they can differ in composition, shape, electric charge distribution, among many other properties. Black holes, on the other hand, only depend on three numbers. This is similar to how a gas can be described by only a few macroscopic parameters (such as temperature, pressure, and volume). 

    In astrophysical scenarios, this description is even simpler. Suppose that a black hole has a very large charge. While gravity is universally attractive, electromagnetism depends on the sign of the charges. As a consequence, the black hole will attract opposite-signed charges much more than it attracts same-signed charges. It follows that the black hole tends to lose charge. Hence, in practical astrophysical scenarios, black holes are usually characterized by their mass and spin, with the electric charge playing little to no role. A stationary black hole of definite mass and spin with no other relevant parameters is known as a Kerr black hole.

    Understanding the properties of the Kerr solution is a daunting task filled with calculations involving differential geometry. Nevertheless, after the calculations are done, some of the results are very simple. For example, general relativity allows us to compute the area of the black hole's event horizon. For a Kerr black hole with mass \(M\) and spin \(J\), the result is given by 
    \begin{equation}\label{eq: area-kerr}
        A = \frac{8 \pi G^2 M^2}{c^4} \qty(1 + \sqrt{1 - \frac{c^2 J^2}{G^2 M^4}}).
    \end{equation}
    Above, \(G\) denotes Newton's constant of gravitation and \(c\) denotes the speed of light. For future use, we notice that we can invert Eq. \eqref{eq: area-kerr} to obtain 
    \begin{equation}\label{eq: kerr-mass-from-area}
        M = \frac{c^2}{\sqrt{16 \pi} G} \sqrt{A + \frac{64 \pi^2 G^2 J^2}{c^6 A}}.
    \end{equation}
    
    Equation \eqref{eq: area-kerr} also tells us that the Kerr black hole has a maximum possible value for its spin---it is not possible to have a black hole unless 
    \begin{equation}\label{eq: max-J}
        J \leq \frac{G M^2}{c}.
    \end{equation}
    Recall that \(J\) is the magnitude of the black hole's spin, and thus \(J \geq 0\) holds automatically. If Eq. \eqref{eq: max-J} is violated, then the area of the event horizon is a complex number, and the object cannot be a black hole. The resulting object would be an example of a naked singularity, which are often seen as unphysical. We will not discuss them here.

\section{Hawking's Area Theorem}\label{sec: hawking-area-theorem}
    By definition, black holes are regions of spacetime from which nothing can escape. As a consequence, it seems reasonable that these regions are only allowed to grow---if a black hole shrinks, it seems something is escaping it. This intuition turns out to be correct. As shown by \textcite{hawking1971GravitationalRadiationColliding,hawking1972BlackHolesGeneral}, within reasonable assumptions, the total area of black holes in the universe must never decrease. In symbols, 
    \begin{equation}\label{eq: area-law}
        \Delta A \geq 0.
    \end{equation}

    Equation \eqref{eq: area-law} may seem familiar. After all, the second law of thermodynamics states that the total entropy of a closed thermodynamical system must never decrease. This is surprisingly similar to the area theorem. Indeed, it is this similarity that first gave rise to the field of black hole thermodynamics, and it is this similarity we will exploit in the following. Notice the analogy goes even further when we consider that, in relativity, the internal energy of a body at rest is given by Einstein's famous formula
    \begin{equation}
        E = M c^2,
    \end{equation}
    with \(M\) being the body's rest mass. Hence, the mass of a black hole is analogous to the internal energy of a gas, while the area of a black hole is analogous to the entropy of a gas. 

\section{Merging Black Holes}
\begin{figure}
        \centering
        \null\hfill\includegraphics{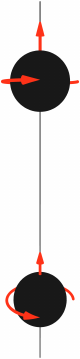}\hfill\includegraphics{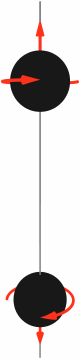}\hfill\null
        \caption{Two black holes spinning around the line that connects them. Their spins can be aligned (as in the left panel) or antialigned (as in the right panel). In this setup, the black holes cannot orbit each other, and may only collide head-on. From a theoretical perspective, this is an interesting scenario because the gravitational waves will not carry away angular momentum, which will simplify our calculations.}
        \label{fig: axisymmetric}
    \end{figure}

    As an interesting example of how this analogy may go, let us consider the following problem. Suppose two Kerr black holes of masses \(M_1\) and \(M_2\) and spins \(J_1\) and \(J_2\). We assume these two black holes are rotating around the line that connects them, so the system has an axis of symmetry. This is illustrated in Fig. \ref{fig: axisymmetric}. We assume the black holes start at a large distance apart, so they can be well approximated as independent objects. Eventually their gravitational attraction will pull them together and they will merge into a single black hole. Once sufficient time has passed, this single black hole will have settled down to a Kerr black hole with mass \(M\) and spin \(J\). 

    Black holes do not merge peacefully. In their collision process, they can deeply disturb spacetime and emit gravitational waves. These are perturbations of spacetime itself that travel at the speed of light, carry energy, and nowadays can be detected at the Earth in gravitational wave observatories. Due to the emission of these waves, the final black hole has less mass than the sum of the initial masses. 
    
    Gravitational waves can also carry angular momentum. This is why we assumed the system to be axisymmetric: this symmetry assumption prevents angular momentum from being radiated away\footnote{A reader experienced with Lagrangian mechanics, or Noether's theorem, may notice that the symmetry of rotation about the \(z\)-axis is related to the conservation of the \(z\)-component of angular momentum. These ideas can also be adapted to general relativity.}. Assuming this is not trivial. Real black holes do not collide in an axisymmetric way, and can orbit around each other several times, for example. In our simplified setup, black holes always collide head-on, and thus their orbital angular momentum vanishes and all their angular momentum is given by the black hole spins. While orbital angular momentum plays an important role in astrophysics, neglecting it here will allow us to learn about black holes with much simpler calculations. 
    
    Even if our assumptions prevent gravitational waves from carrying away angular momentum, they can still carry away energy. A natural question is what is the maximum energy that can be extracted from the system in this way. Notice this question is very similar to asking ``How much work can we extract from a given thermal system?'' because the energy in gravitational waves could in principle be converted into other forms of energy. Just as thermodynamics often worries with heat engines, we can here wonder about ``black hole engines.'' 
    
    To understand this problem, notice the key impositions we must make are the following:
    \begin{enumerate}
        \item The total energy must be conserved. This means, in symbols, 
        \begin{equation}\label{eq: conservation-energy}
            M_1 c^2 + M_2 c^2 = M c^2 + E_{\text{GW}}, 
        \end{equation}
        where \(E_{\text{GW}}\) is the energy emitted in the form of gravitational waves and \(M\) is the mass of the final black hole.
        \item Because we assume the black holes to be rotating around the same line, gravitational waves cannot carry angular momentum away, and all angular momentum is given by the black hole spin. This means angular momentum conservation becomes
        \begin{equation}\label{eq: conservation-angular-momentum}
            \abs{J_1 \pm J_2} = J,
        \end{equation}
        with \(J\) the spin of the final black hole. The positive sign corresponds to the case in which the black hole spins are aligned, and the negative sign corresponds to the case in which the black hole spins are antialigned. See Fig. \ref{fig: axisymmetric} for illustrations. Recall our convention that \(J\) (and \(J_1\) and \(J_2\) as well) are always positive, which is why we add the signs explicitly to account for aligned or antialigned spins.
        \item At last, the area theorem---the total area of black holes in the universe must not decrease. Hence, we have
        \begin{equation}
            A_1 + A_2 \leq A,
        \end{equation}
        with \(A_1\) and \(A_2\) standing for the areas of the initial black holes and \(A\) standing for the area of the resulting black hole.
    \end{enumerate}

    This problem is completely analogous to a ``maximum work problem'' one can find in thermodynamics textbooks---see, for example, the textbook by \textcite{[{}][{, Sec. 4.5.}]{callen1985ThermodynamicsIntroductionThermostatistics}}. From a thermodynamical perspective, we are starting from a state with two black holes, and ending at a state with a single black hole. To maximize the ``work'' (i.e., the emission of gravitational waves) done during this transformation, we should minimize the amount of ``heat'' exchanged during the process. This is achieved in a process in which the total entropy does not change \cite{callen1985ThermodynamicsIntroductionThermostatistics}. In our analogy, this means we should look for a process where the total area does not change. In symbols, we impose \(A = A_1 + A_2\). If the final black hole were any larger, there would be energy stored inside it that could have been emitted instead---at least as far as the area theorem is concerned. Hence, we would be able to emit more gravitational waves by making the black hole smaller. Once \(A = A_1 + A_2\), we have reached the limit and emitted the maximum possible amount of gravitational waves.
    
    With these ideas in mind, we can use conservation of energy (given by Eq. \eqref{eq: conservation-energy}) and Eq. \eqref{eq: kerr-mass-from-area} to find
    \begin{equation}
        E_{\text{GW}} = (M_1 + M_2)c^2 - \frac{c^4}{\sqrt{16 \pi} G} \sqrt{A + \frac{64 \pi^2 G^2 J^2}{c^6 A}}.
    \end{equation}
    Since the maximum possible emission of gravitational waves happens when \(A = A_1 + A_2\), we obtain
    \begin{multline}
        E_{\text{GW}} \leq (M_1 + M_2)c^2 \\ - \frac{c^4}{\sqrt{16 \pi} G} \sqrt{(A_1 + A_2) + \frac{64 \pi^2 G^2 J^2}{c^6 (A_1 + A_2)}}. \label{eq: EGW-Kerr-Efficiency}
    \end{multline}

    Due to conservation of energy, \(E_{\text{GW}}\) must be a fraction of the total available energy, \((M_1+M_2)c^2\). In particular, the larger the initial energy available, the larger the energy that can be emitted in the form of gravitational waves. We thus define
    \begin{equation}\label{eq: gw-emission-efficiency}
        \eta = \frac{E_{\text{GW}}}{(M_1 + M_2)c^2} = 1 - \frac{M}{M_1 + M_2}.
    \end{equation}
    This is the ``efficiency for emission of gravitational waves.'' Just like we can define the thermodynamic efficiency of a heat engine, we can here define the efficiency for emission of gravitational waves of two merging black holes. \(\eta\) is a dimensionless number with \(0 \leq \eta \leq 1\) that measures how much of the initial energy in the black holes was emitted. 
    
    It is interesting to work with \(\eta\) because it is a dimensionless parameter that tells us about how much energy is emitted. We can also introduce some other dimensionless parameters with the goal of making less constants appear in our equations. For example, the difference in the masses of the two colliding black holes is characterized by 
    \begin{equation}\label{eq: delta}
        \delta = \frac{M_1 - M_2}{M_1 + M_2}.
    \end{equation}
    If we convention that \(M_1\) always corresponds to the more massive black hole, then \(M_1 \geq M_2\) and we find that \(0 \leq \delta \leq 1\).

    Using \(c\) and \(G\), we can also obtain dimensionless parameters to measure the spin of the black holes. We can define 
    \begin{equation}\label{eq: dimensionless-spin-parameter}
        \chi_1 = \frac{c J_1}{G M_1^2}\qc\chi_2 = \frac{c J_2}{G M_2^2}, \qq{and} \chi = \frac{c J}{G M^2}.
    \end{equation}
    The initial black holes have the parameters \(\chi_1\) and \(\chi_2\), while the final black hole has \(\chi\). Since \(J\), \(J_1\), and \(J_2\) are the magnitude of their respective black hole spins, they are never negative. Hence, the same is true for \(\chi\), \(\chi_1\), and \(\chi_2\). Furthermore, we know the \(J\)'s must obey Eq. \eqref{eq: max-J}, each for their own black hole. This is translated to \(\chi\) as \(\chi \leq 1\). Therefore, \(\chi\) (and also \(\chi_1\) and \(\chi_2\)) must obey \(0 \leq \chi \leq 1\).

    We can rewrite Eq. \eqref{eq: EGW-Kerr-Efficiency} using the parameters \(\delta\), \(\chi_1\), and \(\chi_2\). To rewrite \(J\) in terms of \(\chi_1\) and \(\chi_2\), we use Eqs. \eqref{eq: conservation-angular-momentum} and \eqref{eq: dimensionless-spin-parameter}. We find the expression
    \begin{widetext}
    \begin{equation}\label{eq: eta-kerr-general}
        \eta \leq 1 - \frac{1}{2\sqrt{2}}\sqrt{(1+\delta)^2\qty(1+\sqrt{1-\chi_1^2})+(1-\delta)^2\qty(1+\sqrt{1-\chi_2^2}) + \frac{\qty((1+\delta)^2\chi_1 \pm (1-\delta)^2 \chi_2)^2}{(1+\delta)^2\qty(1+\sqrt{1-\chi_1^2})+(1-\delta)^2\qty(1+\sqrt{1-\chi_2^2})}},
    \end{equation}
    \end{widetext}
    where the positive signal in \(\pm\) refers to aligned spins, and the negative signal refers to antialigned spins. Given two initial Kerr black holes spinning around the line that connects them, as depicted in Fig. \ref{fig: axisymmetric}, Eq. \eqref{eq: eta-kerr-general} gives the maximum efficiency for emission of gravitational waves allowed by the area theorem. Notice Eq. \eqref{eq: eta-kerr-general} is the statement that the efficiency for emission of gravitational waves is maximized in a process that conserves the black hole area---much like the efficiency of a heat engine is maximized in a process that conserves entropy. To further understand the physical content of Eq. \eqref{eq: eta-kerr-general}, let us analyze a few particular cases.

\section{Limiting Cases}
    \subsection{Vanishing spins}
        First and foremost, let us consider two black holes with vanishing spin (\(J_1 = J_2 = 0\)), but possibly different masses. Nonspinning, uncharged black holes are known as Schwarzschild black holes. In terms of the dimensionless variables, we have \(\chi_1 = \chi_2 = 0\). Equation \eqref{eq: eta-kerr-general} then becomes
        \begin{equation}
            \eta \leq 1 - \sqrt{\frac{1+\delta^2}{2}}.
        \end{equation}

        The maximum efficiency occurs for identical black holes, which means \(\delta = 0\) or \(M_1 = M_2\). In this case, we find
        \begin{equation}\label{eq:efficiency-Sch}
            \eta \leq \frac{2 - \sqrt{2}}{2} \approx 29.3\%.
        \end{equation}
        On the other hand, if one of the two masses vanishes (\(\delta = 1\)), then \(\eta = 0\) and no gravitational waves can be emitted---there is a single black hole, and no merging process takes place. 
    
\begin{figure}[tbp]
            \centering
            \includegraphics{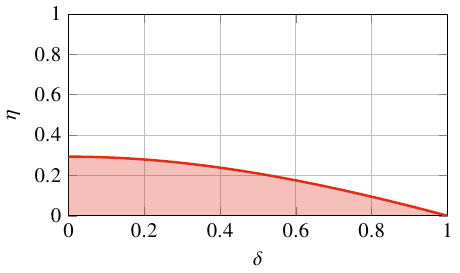}
            \caption{Allowed efficiency \(\eta\) for emission of gravitational waves for a merger of two spinless black holes into a new spinless black hole. The parameter \(\delta\)---see Eq. \eqref{eq: delta}---measures the difference between the masses of the two black holes. Notice more energy can be emitted when the two black holes have similar masses.}
            \label{fig: EGW-efficiency-schwarzschild}
        \end{figure}

        The intermediate cases for two Schwarzschild black holes are shown in Fig. \ref{fig: EGW-efficiency-schwarzschild}.
    
    \subsection{Equal aligned spins}
        As a second case of interest, let us imagine two black holes of equal spins (\(J_1 = J_2\)) which are spinning in the same direction. We still allow them to have different masses. Setting \(\chi_1 = \chi_2 = \chi\) in Eq. \eqref{eq: eta-kerr-general} and picking the upper sign (for aligned spins) yields
        \begin{equation}
            \eta \leq 1 - \sqrt{\frac{1+\delta^2}{2}},
        \end{equation}
        which is the very same result we found when \(\chi = 0\)! If the two black holes have the same spin and orientation, then the efficiency for the emission of gravitational waves does not depend on this spin.

    \subsection{Equal antialigned spins}
\begin{figure}[tbp]
            \centering
            \includegraphics{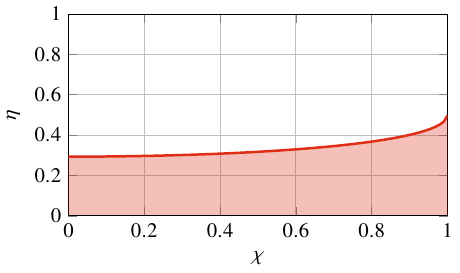}
            \caption{Allowed efficiency \(\eta\) for emission of gravitational waves for a merger of two Kerr black holes with equal masses and equal antialigned spins. \(\chi\) is the magnitude of the dimensionless spin of either one of the black holes. Notice more energy can be emitted when the two black holes rapidly spin in opposite directions.}
            \label{fig: EGW-efficiency-antialigned}
        \end{figure}

        Finally, let us consider two black holes of equal masses (\(M_1 = M_2\), and thus \(\delta = 0\)) and spins of same magnitude (\(J_1 = J_2\)), but spinning in opposite directions. We again set \(\chi_1 = \chi_2 = \chi\) in Eq. \eqref{eq: eta-kerr-general}, but now pick the lower sign (for antialigned spins). After simplification, we find
        \begin{equation}\label{eq: eta-kerr-antialigned}
            \eta \leq 1 - \frac{1}{2} \sqrt{1 + \sqrt{1 - \chi^2}}.
        \end{equation}
        This region is plotted on Fig. \ref{fig: EGW-efficiency-antialigned}.

        In the \(\chi = 0\) case, the black holes are not spinning, and Eq. \eqref{eq: eta-kerr-antialigned} reduces to our previous result for Schwarzschild black holes. It is more interesting to consider the limiting case \(\chi \to 1\), for which the efficiency is maximum. This is the case of maximally rotating black holes, and it yields
        \begin{equation}
            \eta \leq \frac{1}{2} = 50\%,
        \end{equation}
        which is much larger than what we had in the previous cases. 

        Since more energy can be released in the form of gravitational waves when the black holes spin in opposite directions, it seems reasonable they are being attracted to each other more strongly. \textcite{hawking1972BlackHolesGeneral} thus conjectured that general relativity might predict that spinning bodies tend to attract each other when they spin in opposite directions, but repel each other when they spin in the same direction. This was later confirmed by \textcite{wald1972GravitationalSpinInteraction}, who showed there is such a relativistic dipole-like interaction between massive spinning bodies. Notice this is already encoded in the area theorem.

\section{Discussion}
    In the era of gravitational wave observatories, understanding the behavior of these spacetime oscillations is a promising path to uncovering the secrets of the universe. In fact, we may now use gravitational wave observations as tests of general relativity itself.

    In Sec. \ref{sec: hawking-area-theorem} we mentioned Hawking's area theorem holds ``within reasonable assumptions.'' While we will not enter the details of these assumptions, they essentially state that we can neglect quantum effects, that the objects merging are indeed black holes, and that general relativity holds. If a given merger of two massive objects does not obey the area theorem, then one of these hypothesis must be false. This can be tested through gravitational wave observations.

    Take for instance the first direct gravitational wave detection, GW150914 \cite{abbott2016ObservationGravitationalWaves}. This event consisted of the merger of two black holes with masses \(M_1 \approx \SI{36}{\solarmass}\) and \(M_2 \approx \SI{29}{\solarmass}\) into a final black hole of mass \(M \approx \SI{62}{\solarmass}\). Above, \(\si{\solarmass}\) denotes the solar mass. We can estimate the efficiency of emission of gravitational waves in this process as 
    \begin{equation}\label{eq: eta-GW150914}
        \eta_{\text{GW150914}} = 1 - \frac{M}{M_1 + M_2} \approx 4.6\%.
    \end{equation}
    This seems well within the theoretical limitations we could obtain using the area theorem. In fact, for Schwarzschild black holes with these masses, the theoretical limit would be close to \(30\%\), as shown in Eq. \eqref{eq:efficiency-Sch}. This suggests that the data agrees with the area theorem. 
    
    While this calculation works as a sanity check, detailed analyses must take many more variables into account. First and foremost, Eq. \eqref{eq:efficiency-Sch} ignores the spins of the black holes. Nevertheless, even if we did consider them, our discussion throughout the text analyzed only the cases of black holes rotating around the same line, as depicted in Fig. \ref{fig: axisymmetric}. In reality, the spin of each black hole is a three-dimensional vector, and the black holes will be orbiting each other. In addition to the larger number of variables, this also means angular momentum can be carried away by the gravitational waves. Furthermore, the uncertainties in the measured values are very important, since they tell us how sure we can be that the area theorem holds. \textcite{isi2021TestingBlackHoleArea} did such a detailed investigation, and found that the event GW150914 obeyed the area theorem with more than \(95\%\) probability (\(\sim2\sigma\)). We notice that in these tests it is often more convenient to compute the total area of black holes before and after the collision and check that it increases, instead of computing an efficiency bound.

    Much more recently, LIGO observed the GW250114 event \cite{abac2025GW250114TestingHawkings}, which involved black holes of masses \(M_1 \approx \SI{33.6}{\solarmass}\) and \(M_2 \approx \SI{32.2}{\solarmass}\) merging into a final black hole of mass \(M \approx \SI{62.7}{\solarmass}\). Notice the masses are very similar to those in the GW150914 detection. Nonetheless, LIGO has been considerably improved in the ten years between both events, and the uncertainties in GW250114 are considerably smaller. The analysis performed by the LIGO, Virgo, and KAGRA collaborations has found the area theorem was respected with high credibility (\(> 3.4 \sigma\)) in GW250114 \cite{abac2025GW250114TestingHawkings}.

    As the technology of gravitational wave observatories improves---and as new observatories are constructed---the uncertainties in the data will become progressively smaller. This means that, with time, tests of the area theorem will allows us to learn whether the universe indeed respects Hawking's prediction, or whether one of the assumptions behind the theorem is wrong. Therefore, studying whether the area theorem holds in black hole mergers is a way of better understanding how good general relativity is as a description of gravity, and what could lie beyond.

    While computing initial and final areas is more convenient for tests of the area theorem, the efficiency introduced above is a very interesting quantity. Equation \eqref{eq: eta-GW150914} might make it seem that not so much energy is emitted in a black hole merger, since \(5\%\) initially seems to be a small number. That is an incorrect guess. For example, let us estimate the energy emitted by the Sun over its whole existence. We can estimate the energy emitted by the Sun per unit time (its luminosity) as \(L_{\odot} \approx \SI{4e26}{\watt}\). The mass of the Sun can be taken to be \(M_{\odot} \approx \SI{2e30}{\kilo\gram}\), and we will estimate its total lifetime as \(T \approx \SI{1e10}{yr}\). Then the efficiency to convert the solar mass into radiation is 
    \begin{equation}
        \eta_{\odot} = \frac{L_{\odot} T}{M_{\odot} c^2} \approx 0.07\%.
    \end{equation}
    This is such a small number because we are considering the efficiency to convert the rest energy into work. Even the nuclear processes at the Sun have a small efficiency in these terms. Notably, though, black holes are much more efficient, even when they are way within the area theorem bounds. 

    The already large energy emission efficiency of black hole mergers can get even more surprising if we consider a hierarchy of mergers, as first noticed by Misner. If we have two Schwarzschild black holes of equal masses \(M\), they may merge into a final black hole of mass \(2M(1-\eta)\), where \(\eta\) is the efficiency for emission of gravitational waves in a single merger. However, if we have four such black holes, each pair can first merge in a black hole of mass \(2M(1-\eta)\), and these two resulting black holes can merge together into a final black hole of mass \(2^2M(1-\eta)^2\). Again, notice that \(\eta\) is the efficiency \emph{at each merger}. More generally, \(2^N\) initial black holes can describe a hierarchy of mergers into a final black hole of mass \(2^N M(1-\eta)^N\). The total efficiency is then 
    \begin{equation}
        \eta_{\text{tot}} = 1 - \frac{2^N M(1-\eta)^N}{2^N M} = 1 - (1-\eta)^N.
    \end{equation}
    As small as the efficiency \(\eta\) of a single merger may be, large values of \(N\) can bring the total efficiency \(\eta_{\text{tot}}\) very close to \(100\%\).

    Since black hole mergers are already so spectacular, it may seem difficult for remnants of a merger to encounter yet another black hole to merge with. In spite of this, recent observations by the LIGO, Virgo, and KAGRA collaborations report two gravitational wave events in which this may have happened \cite{abac2025GW241011GW241110Exploring}. A black hole that was formed through a merger is expected to have a few distinctive features, such as very large masses compared to their predecessors and a very large spin. The fact that component black holes of the mergers GW241011 and GW241110 present those features suggest they may have been formed in previous coalescences. It cannot be said with certainty that these black holes are indeed of second-generation, but their detection suggests hierarchical mergers could happen in our universe after all. 

    At last, we comment on the analogy between area and entropy. The initial perspective on the area theorem was that it merely mimicked the behavior of entropy. Nevertheless, it was eventually understood that the area of a black hole seems to be intimately related to its entropy in a physical sense. In classical general relativity, this is viewed as absurd because black holes cannot have a nonvanishing temperature. If one were to put a black hole in a box with a gas at temperature \(T > 0\), the system will never be able to reach equilibrium. The black hole absorbs the gas without ever emitting anything in return. The only scenario in which equilibrium could conceivably be achieved would be for \(T = 0\) in a perfect vacuum.

    This perspective, however, changed considerably about fifty years ago, when Hawking \cite{hawking1975ParticleCreationBlack,*hawking1976ParticleCreationBlackErratum} showed that quantum effects can allow a black hole to have a nonvanishing temperature. This supported the view that the area of a black hole should literally be interpreted as its entropy, with the detailed correspondence being given by the Bekenstein--Hawking formula,
    \begin{equation}
        S_{\text{BH}} = \frac{k_B c^3 A}{4G\hbar},
    \end{equation}
    where \(k_B\) denotes Boltzmann's constant and \(\hbar\) denotes the (reduced) Planck constant. This expression brings together ideas from relativity, gravity, thermodynamics, and quantum mechanics into a single result. It is fair to say we still do not understand what it fundamentally means. It is intriguing, though, to notice how the second law of thermodynamics stands between us and our understanding of the inner workings of quantum gravity.

\begin{acknowledgments}
    NAA and CCRE contributed equally to this work and share first authorship.

    Figure \ref{fig: axisymmetric} was drawn using \textsc{Asymptote} \cite{bowman2008AsymptoteVectorGraphics,hammerlindl2004AsymptoteDescriptiveVector}.
 
    We thank Bruno Arderucio Costa for many pleasant and helpful conversations about this work. CCRE also thanks Artur Soares and João Macedo for pedagogical suggestions.

    The work of NAA was supported by the São Paulo Research Foundation (FAPESP) under grant 2025/05161-0. The work of CCRE was supported by the Conselho Nacional de Desenvolvimento Científico e Tecnológico (CNPq) under grant 131663/2025-9.
\end{acknowledgments}
\nocite{*}
\bibliography{bibliography}
\end{document}